\documentclass[11pt,aps,prd,nofootinbib,notitlepage,amsmath,amssymb,floatfix,superscriptaddress]{revtex4}

\usepackage{amsmath}
\usepackage{amssymb}
\usepackage[dvips]{graphicx}
\usepackage{epsfig}
\usepackage{tabularx}
\usepackage{color}

\begin{document}

\title{Hydrodynamic Analogue of Deformed Lorentz Symmetry}
\author{Francesco Marino\footnote{Corresponding Author: \href{mailto:francesco.marino@ino.cnr.it}{francesco.marino@ino.cnr.it}}}
\affiliation{\scriptsize{CNR-Istituto Nazionale di Ottica and INFN, Via Sansone 1, I-50019 Sesto Fiorentino (FI), Italy.}}
\author{Arshid Shabir\footnote{Email: \href{mailto:aslone186@gmail.com}{aslone186@gmail.com}}}
\affiliation{\scriptsize{Canadian Quantum Research Center, 460 Doyle Ave 106, Kelowna, BC V1Y 0C2, Canada.}}
\author{Mir Faizal\footnote{Email: \href{mailto:mirfaizalmir@gmail.com}{mirfaizalmir@gmail.com}}}
\affiliation{ \scriptsize{Irving K. Barber School of Arts and Sciences, University of British Columbia - Okanagan, Kelowna, British Columbia V1V 1V7, Canada.}}
\affiliation{\scriptsize{Canadian Quantum Research Center, 460 Doyle Ave 106, Kelowna, BC V1Y 0C2, Canada.}}
\affiliation{ \scriptsize{Department of Mathematical Sciences, Durham University, Upper Mountjoy, Stockton Road, Durham DH1 3LE, UK.}}
\affiliation{\scriptsize{Computational Mathematics Group, Faculty of Sciences, Hasselt University, Agoralaan Gebouw D, Diepenbeek, 3590, Belgium.}}

\date{\today}

\begin{abstract}
Phonons in a class of viscoelastic fluids have been shown to obey modified dispersion relations analogous to those arising in doubly special relativity (DSR). Here, we extend this analogy by demonstrating that these systems admit a deformed Lorentz symmetry, whose generators, together with translations (and rotations in higher dimensions), define a polynomial deformation of the Poincar\'e algebra. A time-dependent, nonlocal, unitary transformation maps the phonon dynamics to an auxiliary field of Bogoliubov quasiparticles, in which the boost generator and the associated finite transformation can be constructed exactly by integrating the corresponding flow equations. Mapping the symmetry back to the original phonon variables, the corresponding infinitesimal boost generator remains a finite-order local differential operator. While the analogy does not straightforwardly extend to multiparticle sectors it nevertheless provides a concrete physical setting for exploring aspects of DSR phenomenology, such as the inequivalence of kinematical and spectral gamma factors.

\end{abstract}

\maketitle

\section{Introduction}

Doubly Special Relativity (DSR) theories involve a modification of the relativistic energy-momentum relation associated with deformed Lorentz transformations \cite{amelino1,amelino2,amelino3,kowalskiPLB2002,ms}. Besides the light speed, $c$, these theories include an additional observer-independent length scale, $L_p$, usually assumed to be of the order of the Planck length, or equivalently, a maximum energy scale $E_p=\hbar c/L_p$. Although in these theories the Lorentz symmetry is not preserved at the Planck scale, this doesn't imply the loss of covariance and the emergence of a preferred class of inertial observers \cite{amelino2}. The framework is still fully relativistic, but the laws of transformation among observers are deformed in order to render the energy-momentum relation invariant \cite{amelino3,ms}. The deformation becomes negligible in the low-energy limit, so that a DSR model reduces to standard special relativity, exactly as special relativity reduces to Galilei relativity in the low-velocity limit. This scenarios differ from those commonly referred to as Lorentz Invariance Violation (LIV), in which relativistic symmetries break at high energies \cite{mattingly2005,liberati2013}. In LIV frameworks, the transformation laws between inertial observers remain those of special relativity, while the dispersion relations and dynamical laws are not invariant under these transformations, and therefore cannot hold for all observers. 

While various DSR theories have been developed, here we focus on the specific case of the so-called DSR1 model \cite{amelino3}, whose energy-momentum relation expanded up to $\mathcal{O}(E^3/E_p^3)$ is \cite{jafari}
\begin{equation}
E^2 \simeq c^2 {\bf p^2} \left(1+ \frac{E}{E_p} \right) + m^2 c^4
\label{eq1}
\end{equation}
where $E$ is the total energy of a particle with momentum ${\bf p}$ and rest mass $m$.
Eq. \eqref{eq1} includes an energy-dependent correction term which becomes significant when $E \sim E_p$. In the nonrelativistic limit, this deformation can be physically understood as a modification of the particle’s inertial mass, while leaving the rest mass unchanged \cite{jafari}.

While both LIV and DSR dispersion relations naturally arise from attempts to model quantum properties of spacetime in terms of noncommutativity (see e.g. Refs.~\cite{GACandMajid,DimitrijevicJHEP2011}) or discreteness (see e.g. Refs.~\cite{urrutia, hanno}), modified dispersion relations with an emergent low-energy Lorentz symmetry are also a generic feature of collective excitations in classical and quantum fluids \cite{agrev,natrev}. Notable examples include the dispersion relation of surface gravity waves in shallow water \cite{rousseaux}, as well as the Bogoliubov dispersion relation in atomic \cite{bogoliubov,reviewBEC} and photonic Bose gases \cite{vocke15,fontaine18,falque}.
In these systems, the surface wave velocity and the sound speed respectively act as analogues of the invariant speed of light in special relativity, while the scale at which the effective Lorentz symmetry is broken plays a role analogous to the Planck scale \cite{natrev}. 

Although all the hydrodynamic systems mentioned above (and, more generally, all condensed-matter mean-field models) ultimately exhibit an explicit breaking of Lorentz symmetry at sufficiently small scales, where the underlying microscopic physics becomes relevant and the mean-field description ceases to be valid (e.g., the mean intermolecular distance in classical fluids or the healing length in Bose–Einstein condensates), it is natural to ask whether one can construct physically consistent models in which Lorentz transformations are instead deformed so as to leave both the modified dispersion relation and the associated propagation equations invariant, in close analogy with DSR scenarios.

In a recent work, it was shown that elementary excitations (phonons) in a class of viscoelastic fluids can mimic aspects of DSR kinematics \cite{marino25}. The analogy was established at the level of effective Klein–Gordon equations and the associated modified dispersion relations. In particular, depending on the nature of the viscoelastic interactions, phonons can reproduce dispersion relations analogous to those arising in both DSR1 \cite{amelino3} and DSR2 \cite{ms} models, with an analogue Planck scale emerging from the non-Newtonian behavior at high phonon energies.

In this work, we extend this correspondence by constructing an explicit deformed boost generator for the physical phonon field that exactly preserves its evolution equation and, together with translations and rotations, closes in arbitrary spatial dimension into a nonlinear (polynomial) algebra, whose explicit form and finite transformations are worked out in detail in (1+1) dimensions.

By performing a time-dependent transformation that diagonalizes the dynamics, the phonon field can be mapped onto a Bogoliubov quasiparticle field obeying a modified Klein–Gordon equation. In this representation, the underlying deformed Lorentz symmetry becomes manifest, with boost transformations acting locally in both momentum and real space. Mapping back to the original field then yields the corresponding symmetry transformations for the physical phonons. While the analogy no longer holds for multiparticle states, since phonon momenta add linearly in contrast with the nonlinear composition laws typically associated with deformed boosts, it nevertheless provides a concrete physical framework in which to explore phenomenological effects governed by the deformation scale. One such effect is the recently studied modification of time dilation and the emergence of non-equivalent "gamma factors" in DSR models  \cite{AmelinoCamelia2025TimeDilation}, a feature that could, in principle, discriminate between LIV and DSR scenarios.

\section{Acoustic analogue model}

We briefly recall the results of Ref. \cite{marino25}, showing that elementary fluctuations (phonons) in a class of barotropic, viscoelastic fluids obey a DSR1-modified Klein-Gordon equation.

We start from the Navier-Stokes equations for mass and momentum conservation
\begin{subequations} \label{eq2}
\begin{eqnarray}
\partial_t \rho + \nabla\cdot(\rho {\bf v}) = 0 \label{eq2a} \, \\ 
\rho [\partial_t {\bf v} + ({\bf v} \cdot {\bf \nabla}) {\bf v}]= - {\bf \nabla} P - \rho {\bf \nabla} f \,+ \, \rho {\bf \nabla} \Phi  \label{eq2b} 
\end{eqnarray}
\end{subequations}
where $\rho$ is the mass density and $\mathbf{v}$ the velocity field. The three force terms on the right-hand side of Eq.~\eqref{eq2b} each play a distinct physical role. The first, $-\nabla P$, is the pressure gradient arising from a generic barotropic equation of state $P = P(\rho)$, in which the pressure depends solely on the local density. The second, $-\nabla f$, derives from a repulsive dipolar potential $\phi$ satisfying the Poisson equation $\nabla^2  f = -\beta\rho$ with $\beta > 0$, constant. The third, $\nabla\Phi$, encodes viscoelastic forces through the phenomenological relation $\nabla\Phi = i\eta\nabla^2\mathbf{v}$, with $\eta>0$ constant. In contrast to conventional viscosity, $\nu \nabla^2 \mathbf{v}$ ($\nu$ real), associated with momentum dissipation \cite{landau}, the term $\nabla \Phi$ describes a reactive, non-dissipative contribution. In viscoelastic fluids, applied stresses do not solely dissipate energy, but also temporarily store it in the deformation of the internal microstructure, which can subsequently be released to restore the equilibrium configuration. In the frequency domain (with the Fourier convention $\exp(-i\omega t)$
used below), the viscoelastic response is described by the complex
modulus $G^*(\omega)=G'(\omega)-iG''(\omega)$, where $G'$ is the storage
modulus and $G''$ is the loss modulus \cite{chhabra,irgens,ferry}. The
corresponding complex kinematic viscosity is $\eta^*(\omega)=i
\frac{G^*(\omega)}{\rho \omega}$, so that the storage modulus gives the
reactive, phase-shifted component, while $G''(\omega)$ contributes to
the dissipative viscous component. A constant imaginary coefficient $i\eta$ corresponds to a storage modulus $G'(\omega) \simeq \rho \eta \omega$
over the (positive) frequency band of interest, with the loss modulus
$G''(\omega)\ll G'(\omega)$ (i.e. dissipation assumed small). Power-law
viscoelastic responses, as in critical gels \cite{winter}, provide
examples of smoothly frequency-dependent moduli that admit local linear
approximations over a limited band, while neglecting the dissipation
restricts the description to phonon modes and observation times for which
damping remains small. We stress that this is a phenomenological, band-limited description of the reactive response for positive-frequency phonon modes, and is not derived from a specific causal memory kernel or
microscopic model of the fluid.

The real coefficient $\eta$ carries units of $[\text{length}]^2/[\text{time}]$ and defines a characteristic length scale $L_p = \eta/c_s$ at which the viscoelastic and bulk pressure effects become comparable \cite{marino25}. As we shall see, this scale plays the role of the analogue Planck length in the emergent DSR kinematics.

We now derive the phonon propagation equations in this fluid. Phonons are defined as first-order fluctuations $\{\rho_1, {\bf v_1}\}$ of the quantities describing the background fluid flow $\{\rho_0, {\bf v_0}\}$. We thus decompose the the total density and velocity fields as $\rho = \rho_0 + \rho_1$ and ${\bf v} = {\bf v_0} + {\bf v_1}$, with $\rho_1 \ll \rho_0$ and $\lVert \mathbf{v_1} \rVert \ll \lVert \mathbf{v_0} \rVert$.
Since here we are not interested in phenomena related to curved acoustic spacetime, we restrict ourselves to a homogeneous background, assuming constant density $\rho_0$ and constant flow $\mathbf{v}_0$. We further assume that the flow is irrotational, $\nabla\times\mathbf{v} = 0$, so that the velocity field can be written as the gradient of a scalar potential, $\mathbf{v} = \nabla\psi$, such that ${\bf v_0}={\bf \nabla} \psi_0$ and ${\bf v_1}={\bf \nabla} \psi_1$. Substituting these expressions into Eqs.~\eqref{eq2}, we obtain the linearized continuity and Euler equations
\begin{subequations} \label{eq3}
\begin{eqnarray}
\partial_T \rho_1 + \rho_0 \nabla^2 \psi_1 = 0 \label{eq3a}\; \\
\partial_T \psi_1 = - \frac{P_1}{\rho_0} -  f_1  \, + \, \Phi_1 \label{eq3b} 
\end{eqnarray}
\end{subequations}
where $P_1$, $f_1$, and $\Phi_1$ denote the first-order fluctuations of the bulk pressure, the dipolar potential, and the viscoelastic potential, respectively, and $\partial_T = \partial_t + \mathbf{v}_0 \cdot \nabla$ is the comoving (material) derivative. The coupled system~\eqref{eq3} fully determines the phonon dynamics. 

We now evaluate the three terms on the right-hand side of Eq.~\eqref{eq3b}. Expanding the bulk pressure as $P(\rho_0 + \rho_1) \approx P(\rho_0) + \partial_\rho P(\rho_0) \,  \rho_1$, we obtain $P_1 = c_s^2 \rho_1$, where $c_s^2 = \partial_\rho P(\rho_0)$ defines the sound speed. In the analogue gravity framework, $c_s$ plays the role of the invariant speed of light in special relativity \cite{agrev}.
The fluctuations in the dipolar potential $f_1$ satisfy $\nabla^2 f_1 = -\beta \rho_1$. Applying $\partial_T$ to this equation and using Eq.~\eqref{eq3a} gives $\nabla^2\left(\partial_T f_1-\beta\rho_0\psi_1\right)=0$. In the confined-fluid configuration considered here, we impose homogeneous Neumann boundary conditions. The associated elliptic problem determines the solution up to an additive spatial constant. Since this zero mode does not contribute to the phonon dynamics, it can be set to zero, yielding $\partial_T f_1=\beta\rho_0\psi_1$. Finally, the viscoelastic contribution is given by $\Phi_1 = i \eta \nabla^2 \psi_1$.

We now combine these results to obtain a closed wave equation for $\psi_1$. Applying the material derivative $\partial_T$ to both sides of the linearized Euler equation~\eqref{eq3b} and using the expressions above together with Eq.~\eqref{eq3a}, we obtain 
\begin{equation}
\partial_{T}^2 \psi_1- c_s^2 \nabla^2 \psi_1 + \Omega_0^2 \psi_1  \, - \, i \eta \partial_T \nabla^2 \psi_1 =0 \, ,
\label{eq4}
\end{equation}
where $\Omega_0=\sqrt{\beta \rho_0}$ is the phonon rest frequency. This equation reveals the essential role of the dipolar interaction: for standard barotropic fluids in which $\beta =\eta = 0$, phonons are massless excitations. The inclusion of repulsive dipolar interactions introduces a frequency gap $\Omega_0$ below which propagating modes cannot exist, precisely analogous to the rest-mass gap $M c^2/\hbar$ of a massive relativistic particle \cite{marino2019,ciszak2021,marino2024}.

The connection between Eq. \eqref{eq4} and the DSR1 energy-momentum relation~\eqref{eq1} is established by the standard quantum-mechanical identifications $E = i\hbar\,\partial_T$ and $\mathbf{p} = -i\hbar\nabla$. Setting $m = \hbar\Omega_0/c_s^2$, and introducing the analogue Planck energy $E_p = \hbar c_s^2 / \eta=\hbar c_s / L_p$, Eq.~\eqref{eq4} yields 
\begin{equation}
E^2 = c_s^2 {\bf p^2} \left(1+ \frac{E}{E_p} \right) + m^2 c_s^4 \, ,
\label{eq:DSR1recovered}
\end{equation}
which is precisely the DSR1 relation~\eqref{eq1} with the replacements $c \to c_s$. The analogue Planck energy $E_p$ is the energy scale at which the viscoelastic correction becomes comparable to the leading acoustic term: when $E \sim E_p$, the factor $(1 + E/E_p)$ departs significantly from unity and the deformation of the relativistic kinematics cannot be neglected. The hierarchy $L_p \ll \lambdabar_C$, where $\lambdabar_C= c_s/\Omega_0$ is the phonon reduced Compton wavelength, ensures a wide regime of "acoustic relativity" in which the kinematics are approximately Lorentzian before the DSR deformation sets in, mirroring the expected hierarchy between the Planck scale and particle rest masses in quantum gravity scenarios.

For simplicity, we rescale the time coordinate by the phonon rest frequency $T \rightarrow \Omega_0 T$ and the spatial coordinates by $\lambdabar_C$, ${\bf x} \rightarrow {\bf x}/\lambdabar_C$. In terms of these rescaled coordinates, Eq. \eqref{eq4} takes the form of a deformed Klein-Gordon equation
\begin{equation}
\partial_{T}^2 \psi_1 -  \nabla^2 \psi_1 + \psi_1  \, - \, i \epsilon \partial_T \nabla^2 \psi_1=0 \, ,
\label{eq5}
\end{equation}
where the dimensionless parameter $\epsilon = \hbar \Omega_0 / E_p = L_p/\lambdabar_C$ quantifies the deformation.
Eq. \eqref{eq5}  is a classical linear wave equation for the
complex positive-frequency component of the perturbation; its plane-wave solutions determine the dispersion relation.
The energy and momentum notation introduced above serves only to express the formal analogy with DSR kinematics. Solving Eq. \eqref{eq:DSR1recovered} for $E$ yields two roots, which are not related by $E\to -E$ because of the presence of the deformation.
The physical phonon modes considered in this work are described by the positive root, $E_+(p)$, since the reactive term is introduced as a phenomenological description valid over a positive-frequency band (see the discussion after Eqs.~\eqref{eq2}). The negative-frequency component of the real disturbance is the complex conjugate of the positive-frequency one and obeys the conjugate equation, with energies $E=-E_+(p)$. The second root of Eq.~\eqref{eq:DSR1recovered} is therefore not used to describe the real disturbance, and we do not assign any physical interpretation to it. In particular, no antiphonon interpretation is assumed, and the unequal roots alone do not establish a physical discrete-symmetry violation, which would require a specified quantization of the phonon field.


\section{Deformed Boost generator}

The deformed Klein-Gordon equation \eqref{eq5} can be written compactly as
\begin{equation}
\mathcal{L} \psi=0 \, 
\label{eq6}
\end{equation}
where we introduced the linear differential operator $\mathcal{L} = \Box \,- \, i\epsilon\,\partial_T\nabla^2$ and, for simplicity of notation, the new variable $\psi \equiv \psi_1$.

In the undeformed case ($\epsilon = 0$), the operator $\mathcal{L}$ reduces to the massive d'Alembertian $\Box = \partial_T^2-\nabla^2 + 1$, which is invariant under the generator of standard Lorentz boosts along the $a$-th spatial direction, $K_a = T\partial_{x_a} + x_a\partial_T$, ($a = 1,2,3$). When $\epsilon \neq 0$, however,  the term $i\epsilon\,\partial_T\nabla^2$ transforms inhomogeneously under $K_a$ because the boost mixes time and space derivatives, and the resulting expression does not reduce to a multiple of $\mathcal{L}$. The evolution is thus non-diagonal, which makes the identification of the underlying symmetry and the construction of a generalized boost generator more involved. 

We therefore introduce the auxiliary field $\phi = U^{-1}\psi$, where $U = e^{\frac{i}{2}\epsilon T \nabla^2}$ is a time-dependent unitary transformation \footnote{The operator $U$ is unitary in the $L^2(\mathbb{R}^d)$ sense, $U^\dagger = U^{-1} = e^{-\frac{i}{2}\epsilon T\nabla^2}$, since $\nabla^2$ is self-adjoint (with appropriate domain) and the exponent is purely imaginary.}. In position space, $U$ involves the Laplacian $\nabla^2$ to all powers, and therefore acts as an infinite-order nonlocal differential operator. In momentum space, by contrast, $\tilde{U} \to e^{-\frac{i}{2}\epsilon T k^2}$, which acts locally on each Fourier mode. Physically, $U$ can be recognized as the time-evolution operator generated by the free-particle Schr\"odinger Hamiltonian $H = -\frac{\epsilon}{2}\nabla^2$, with $T$ playing the role of the evolution parameter.
Substituting $\psi = U \phi$ into Eq. \eqref{eq6}, and multiplying from the left by $U^{-1}$ we obtain
\begin{equation}
\mathcal{L}_\phi \phi \equiv U^{-1} \mathcal{L} U \phi = 0  \, .
\label{eq7a}
\end{equation}
Since $\nabla^2$ commutes with $U$ and $U^{-1}$, and $U^{-1} \partial_T U = \partial_T + \tfrac{i}{2}\epsilon \nabla^2$, a direct computation yields $\mathcal{L}_\phi = \Box + \tfrac{\epsilon^2}{4} \nabla^4$ and the equation of motion \eqref{eq6} takes the form
\begin{equation}
\left( \Box  + \tfrac{\epsilon^2}{4} \nabla^4 \right) \phi = 0 \, .
\label{eq7}
\end{equation}

In terms of the auxiliary field $\phi$, the dynamics is diagonal, in the sense that the temporal and spatial evolution of the field are decoupled. Passing to Fourier space with the convention $\partial_T \to -i\omega$ and $\nabla \to i\mathbf{k}$, Eq. \eqref{eq7} leads to the dispersion relation
\begin{equation}
\omega^2 = 1 + k^2 \left(1 + \frac{\epsilon^2}{4} k^2 \right) \, ,
\label{bogo}
\end{equation}
where $k=\vert {\bf k} \vert$. This corresponds to a massive generalization of the Bogoliubov dispersion relation describing fluctuations in weakly interacting Bose gases \cite{marino2019,reviewBEC}. The nonlocal transformation $U$ thus maps the phonon field of the classical viscoelastic fluid \eqref{eq2} into a field of elementary excitations analogous to those found in certain quantum fluids. Similar dispersion relations arise, for instance, in coupled condensates \cite{silke-visser} (see also the recent experiment \cite{ferrari22}) and in superfluid systems with suitably engineered interactions \cite{eg2,marino2019,marino2024}.

We emphasize that the physical phonon field, i.e. the one that would be probed in an experimental realization of the viscoelastic fluid, is precisely described by the fluctuations variable $\psi$, whose equation of motion follows directly from the linearized hydrodynamic equations \eqref{eq3}.
The field $\phi$ provides only an auxiliary variable chosen to diagonalize the dynamics.  The local quartic Klein--Gordon equation obtained for $\phi$ should therefore be understood as an equivalent representation of the physical phonon dynamics, rather than as a separate physical theory.

A linear differential operator $D_a$ is a symmetry generator of the equation $\mathcal{L}_\phi\,\phi = 0$ if and only if it maps every solution to another solution. This is guaranteed by the condition 
\begin{equation}
[D_a,\mathcal{L}_\phi]=0 \, .
\label{cond}
\end{equation}
We impose this condition on the class of finite-order, rotationally covariant vector differential operators that are at most linear in the coordinates and whose spatial derivative structure is built from powers of $\nabla^2$. This class is sufficiently general to contain the ordinary boost $K_a$ and its rotationally covariant deformations compatible with the derivative structure of the theory. The most general operator in this class is
\begin{equation}
D_a= T\partial_{x_a}F(\nabla^2)+x_a\partial_TG(\nabla^2)
+\partial_{x_a}R(\nabla^2),
\end{equation}
where $F$, $G$, and $R$ are polynomials in $\nabla^2$, with coefficients analytic in $\epsilon^2$. Substitution into the symmetry condition \eqref{cond} gives
\begin{equation}
[D_a,\mathcal{L}_\phi]
=2\partial_{x_a}\partial_T
\left[\left(1-\frac{\epsilon^2}{2}\nabla^2\right)G(\nabla^2)-F(\nabla^2)\right].
\end{equation}
Therefore, imposing the symmetry condition \eqref{cond} requires
\begin{equation}
F(\nabla^2)=
\left(1-\frac{\epsilon^2}{2}\nabla^2\right)G(\nabla^2).
\end{equation}
There remains a freedom in the choice of $G(\nabla^2)$, as well as the possibility of adding the separately commuting term $\partial_{x_a}R(\nabla^2)$ because it commutes with $\mathcal{L}_\phi$. 
We choose $G(\nabla^2)=1$ and $R(\nabla^2)=0$, which define the minimal canonical representative and ensure the standard normalization and continuous reduction to the usual Lorentz boost as $\epsilon\to0$. The resulting deformed boost is therefore
\begin{equation}
D_a=K_a-\frac{\epsilon^2}{2}T\partial_{x_a}\nabla^2 \, .
\label{eq9}
\end{equation}

A nontrivial feature of the present model is that the deformed Lorentz symmetry admits an exact realization in terms of a local, finite-order differential operator. This contrasts with many DSR-inspired scenarios, where nonlinear realizations of Lorentz symmetry in momentum space generically produce generators that, when represented in position space, involve infinite series of derivatives and are therefore nonlocal. Locality in the present model is a nontrivial consequence of the deformation structure: the quartic spatial term in $\mathcal{L}_\phi$ is exactly compensated by the third-order correction to the boost generator. The resulting commutators remain polynomial in the basic generators, but generate higher-order polynomial combinations under iteration. The symmetry is therefore naturally described as a nonlinear polynomial deformation rather than as an ordinary finite-dimensional Lie algebra. In this framework, the quartic dispersion \eqref{bogo} lies at the boundary between standard relativistic theories and fully nonlocal DSR models, allowing for a local infinitesimal implementation of deformed Lorentz symmetry at the single-particle level. This characteristic is not expected to extend to more general classes of dispersion relations.

With the previous Fourier convention, the corresponding operator identifications are, ${\bf k}=-i {\bf \nabla}$, $\omega=i\partial_T$, $x_a=i\partial_{k_a}$ and $T=-i\partial_\omega$. We therefore obtain the deformed boost generator in Fourier space $(\omega, {\bf k})$
\begin{equation}
\tilde{D}_a = \omega \partial_{k_a} + k_a \left(1 + \frac{\epsilon^2}{2} k^2 \right) \partial_\omega \, ,
\label{eq10}
\end{equation} 
under which the dispersion relation \eqref{bogo} is invariant. For $\epsilon=0$, Eq.~\eqref{eq10} reduces to the standard Lorentz boost $\omega \partial_{k_a} + k_a \partial_\omega$, consistent with ordinary special relativity. 

The corresponding boost generator for the physical phonon field $\psi$ is obtained by mapping $D_a$ back through the inverse transformation, yielding the operator $D_a^\psi = U D_a U^{-1}$. Despite the nonlocal character of the field transformation $U$, the boost generator acting on the physical phonon field remains local and contains only finitely many derivatives.
The Baker--Campbell--Hausdorff series terminates after the second commutator, yielding the finite-order local differential operator
\begin{equation}
D_a^\psi=T\partial_{x_a}+x_a\partial_T+i\epsilon T\partial_{x_a}\partial_T-\frac{i\epsilon}{2}x_a\nabla^2 \, .
\end{equation}
The two descriptions in terms of the physical phonon field $\psi$ and the auxiliary field $\phi$ are thus equivalent representations of the same deformed symmetry. In both representations the equations of motion are local and the infinitesimal boost generators are finite-order local differential operators. Thus, in the present construction, the nonlocality resides only in the field redefinition relating the two representations.

\subsection{Deformed Poincar\'e algebra}

In one spatial dimension, the boost generator \eqref{eq9} can be combined with translations $H = i \partial_T$, $P = -i\partial_x$ to define a deformed symmetry algebra
\begin{eqnarray}
[H,P] = 0 \, , \qquad [D,P] = i \partial_T = H \, \\
{[D,H]} = -i \partial_x + \frac{\epsilon^2}{2} i \partial_x^3= P + \frac{\epsilon^2}{2} P^3
\end{eqnarray}
which reduces to the standard Poincar\'e algebra in the limit $\epsilon \to 0$. The deformation resides entirely in the commutator $[D,H]$, where the dependence on $P$ is cubic rather than linear. This structure is reminiscent of $\kappa$-Poincar\'e algebras \cite{luki,luki2,kowalskiPLB2002}, which also features nonlinear deformations of the boost-momentum commutator, although the specific functional form differs. In contrast, the relation $[H,P]=0$ remains unchanged, indicating that spacetime homogeneity is preserved and translations are not affected by the deformation. 

The commutators can be expressed polynomially in the basic generators $\{D,P,H\}$, although iterated commutation belongs to the associated polynomial enveloping algebra rather than to an ordinary finite-dimensional Lie algebra; for example, $[D,P^3]=3P^2H$. Remarkably, in $(1+1)$ dimensions (or for vanishing transverse momentum) this realization allows us to integrate the finite transformations explicitly and investigate their consequences for time dilation, an effect that has been recently investigated in the DSR literature \cite{AmelinoCamelia2025TimeDilation}.

In higher spatial dimensions, the differential operator $D_a$ in Eq. (15) continues to define a symmetry of the auxiliary-field equation. The algebraic structure, however, requires the inclusion of the rotation generators $J_{ab}=x_a\partial_b-x_b\partial_a$. 
The relevant commutators are
\begin{equation}
[D_a,D_b]=\left(1+\frac{\epsilon^2}{2}\mathbf P^2\right)J_{ab}, \qquad [J_{ab},D_c] = \delta_{bc}D_a-\delta_{ac}D_b \nonumber
\end{equation}
together with
\begin{equation}
[D_a,P_b]=\delta_{ab}H, \qquad [D_a,H]= P_a\left(1+\frac{\epsilon^2}{2}\mathbf P^2\right) .
\end{equation}
Thus, the deformed boosts, rotations, and translations form a polynomially closed algebra. The closure is not that of an ordinary finite-dimensional Lie algebra with constant structure coefficients, since successive commutators generate polynomial functions of the translation generators. The $(1+1)$-dimensional case is special because there is only one boost and no independent spatial rotation generator, allowing the algebra to be expressed using the smaller set $\{H,P,D\}$.

\subsection{Finite boost transformation}

The finite form of the deformed boost is most conveniently constructed in Fourier space, where the symmetry acts as a flow on the on-shell surface defined by Eq.~\eqref{bogo}. Starting from the generator~\eqref{eq10}, the finite transformation with rapidity parameter $\lambda$ is obtained by integrating the flow equations
\begin{subequations} \label{eqb}
\begin{eqnarray}
\frac{dk_a}{d\lambda} &=& \omega \label{eqb1} \, \\
\frac{d\omega}{d\lambda} &=& k_a \left(1 + \frac{\epsilon^2}{2}k^2\right) \label{eqb2}\, ,
\end{eqnarray}
\end{subequations}
Eqs. \eqref{eqb} have the structure of ordinary Lorentz boost flow equations, except that the second equation is modified by the momentum-dependent factor $(1 + \frac{\epsilon^2}{2}k^2)$. 

As we shall see, these equations admit exact analytic solutions in terms of Jacobi elliptic functions. It is nonetheless instructive to first present the $\mathcal{O}(\epsilon^2)$ solution obtained via the Poincar\'e–Lindstedt method \cite{strogatz}, as it illustrates how the deformed symmetry emerges perturbatively from the standard Lorentz transformation. In $(1+1)$ dimensions, $k = k_a$, and differentiating Eq.~\eqref{eqb1} with respect to $\lambda$, and using Eq.~\eqref{eqb2}, we obtain
\begin{equation}
\frac{d^2k}{d\lambda^2} = k \left(1 + \frac{\epsilon^2}{2}k^2\right) \label{eq11}\, ,
\end{equation}
with initial conditions $k(0)=k_0$ and $d k(0)/d\lambda=\omega_0$. Eq. \eqref{eq11} is a Duffing-type equation without external forcing, a paradigmatic example for which the above perturbative method applies \cite{strogatz}. We therefore introduce the $\mathcal{O}(\epsilon^2)$ expansion of the solution $k(\Lambda)=k^{(0)} (\Lambda)+ \epsilon^2 k^{(1)}(\Lambda) + \mathcal{O}(\epsilon^4)$ and the deformed rapidity $\Lambda= \lambda (1+\alpha \epsilon^2 + \mathcal{O}(\epsilon^4) )$ necessary to remove the resonant (secular) terms. Substituting these expressions into Eq.~\eqref{eq11} and truncating consistently at order $\mathcal{O}(\epsilon^2)$, the requirement of eliminating secular growth fixes
\begin{equation}
\alpha = \frac{3}{16} (k_0^2 - \omega_0^2) \, .
\end{equation}
The resulting order-$\epsilon^2$ approximation for the boosted momentum reads
\begin{eqnarray}
k'(\lambda)&=&k_{0}\cosh\Lambda+\omega_{0}\sinh\Lambda
 \nonumber \\ &+&\epsilon^{2}\Big[
c_{1}k_{0}\big(\cosh 3\Lambda-\cosh\Lambda\big)
+c_{2}\omega_{0}\sinh 3\Lambda
-\big(\alpha+3c_{2}\big)\omega_{0}\sinh\Lambda
\Big]
+\mathcal{O}(\epsilon^{4}),
\label{eq12}
\end{eqnarray}
where $c_{1}=(k_{0}^{2}+3\omega_{0}^{2})/64$, 
$c_{2}=(3k_{0}^{2}+\omega_{0}^{2})/64$.
\label{eqPL7} 
Differentiating with respect to $\lambda$, one obtains the corresponding boosted frequency
\begin{eqnarray}
\omega'(\lambda)&=&\omega_{0}\cosh\Lambda+k_{0}\sinh\Lambda
  \nonumber \\ &+&\epsilon^{2}  \Big[
(\alpha-c_{1})k_{0}\sinh\Lambda
-3c_{2}\omega_{0}\cosh\Lambda
+3c_{1}k_{0}\sinh 3\Lambda
+3c_{2}\omega_{0}\cosh 3\Lambda
\Big]
+\mathcal{O}(\epsilon^{4}).
\label{eq13}
\end{eqnarray}

Eqs.~\eqref{eq12}–\eqref{eq13} represent a nonlinear deformation of Lorentz boosts. Notably, the zeroth-order contribution retains the exact structure of a standard Lorentz transformation, with the original rapidity $\lambda$ replaced by the renormalized parameter $\Lambda$. At this order, the deformation manifests solely through a modification of the rapidity, without altering the functional form of the transformation. At $O(\epsilon^2)$, however, additional corrections arise in the form of higher-order hyperbolic contributions (e.g.\ $\cosh 3\Lambda$, $\sinh 3\Lambda$), which are analogous to the third-harmonic components found in the Poincar\'e–Lindstedt solution of the Duffing anharmonic oscillator 
\footnote{The Duffing oscillator equation is simply obtained from Eq.~\eqref{eq11} upon reversing the sign of the right-hand side.}. In the limit $\epsilon \to 0$, one recovers $\Lambda \to \lambda$ and all higher-order contributions vanish, reducing the transformation to the standard Lorentz boost.

The exact solution of Eqs. \eqref{eqb} can be obtained by noting that on the positive-frequency branch, $\omega = \sqrt{1 + k^{2} + \tfrac{\epsilon^{2}}{4}k^{4}} \equiv \omega_{\phi}(k)$ obtained by solving Eq. \eqref{bogo} the first flow equation reduces to a separable first-order ODE in $k$ alone:
\begin{equation}
    \frac{dk}{d\lambda}
    = \omega_{\phi}(k)
    = \sqrt{1 + k^{2} + \frac{\epsilon^{2}}{4}k^{4}},
\label{eq:sep_ode}
\end{equation}
with the formal integral
\begin{equation}
    \lambda
    = \int_{k_{0}}^{k'(\lambda)}
      \frac{dk}{\sqrt{1 + k^{2} + \dfrac{\epsilon^{2}}{4}k^{4}}}= \frac{1}{\mu}[F\!\left(\arctan(k'/\sqrt{\alpha_{-}}),\kappa\right)-u_{0}] \, ,
\label{eq:integral}
\end{equation}
where $F(\varphi,\kappa) = \int_{0}^{\varphi} d\theta/\!\sqrt{1-\kappa^{2}\sin^{2}\theta}$ is the incomplete elliptic integral of the first kind, $\alpha_{\pm}$ are the positive factorization parameters
\begin{equation}
    \alpha_{\pm}
    \;=\; \frac{2}{\epsilon^{2}}\!\left(1 \pm \sqrt{1-\epsilon^{2}}\right)
    \;>\; 0,
    \qquad \alpha_{+} > \alpha_{-} > 0 \, ,\qquad 0<\epsilon<1 ,
\label{eq:alphas}
\end{equation}
such that  
\begin{equation}
 1+k^2+\frac{\epsilon^2}{4}k^4
 =\frac{\epsilon^2}{4}(k^2+\alpha_-)(k^2+\alpha_+)\, ,
\label{eq:factorization}
\end{equation} 
$\mu=\epsilon \sqrt{\alpha_+}/2$, $\kappa=\sqrt{1-\alpha_-/\alpha_+}$ and $u_{0}= F(\arctan k_{0}/\sqrt{\alpha_{-}},\,\kappa )$. 

Inverting by the Jacobi amplitude denoted by $\mathrm{am}(u,\kappa)$, so that $\arctan(k'/\sqrt{\alpha_{-}}) = \mathrm{am}(u_{0}+\mu\lambda,\kappa)$, and using the identity $\tan(\mathrm{am}(u,\kappa)) = \mathrm{sc}(u,\kappa) = \mathrm{sn}(u,\kappa)/\mathrm{cn}(u,\kappa)$, where $\mathrm{sn}(u,\kappa)$ and $\mathrm{cn}(u,\kappa)$ are the Jacobi elliptic sine and cosine functions, respectively, we obtain the \emph{exact} finite deformed boost for the auxiliary field:
\begin{align}
    k'_{a}(\lambda)
    &= \sqrt{\alpha_{-}}\;
       \mathrm{sc}(u_{0} + \mu\lambda,\;\kappa),
    \label{eq:finiteboost_k} \\[4pt]
    \omega'(\lambda)
    &= \omega_{\phi}(k'_{a}(\lambda))
    = \sqrt{1 + k_{a}^{\prime\,2}
             + \frac{\epsilon^{2}}{4}k_{a}^{\prime\,4}} \, .
    \label{eq:finiteboost_omega}
\end{align}
For $\epsilon \to 0$, one has $\alpha_{-} \to 1$, $\mu \to 1$, and $\kappa \to 1$. In the degenerate limit $\kappa = 1$ the Jacobi elliptic functions reduce to hyperbolic functions: $\mathrm{sc}(u,1) = \sinh u$. Furthermore, $u_{0} \to F(\arctan k_{0},\,1) = \operatorname{arcsinh}(k_{0})$, using the identity $F(\varphi,1) = \operatorname{arcsinh}(\tan\varphi)$. Hence
\begin{align}
    k'(\lambda)
    &\;\xrightarrow{\;\epsilon\to 0\;}
    \sinh\!\bigl(\operatorname{arcsinh}(k_{0})+\lambda\bigr)
    = k_{0}\cosh\lambda + \omega_{0}\sinh\lambda,
    \label{eq:boost_limit_k} \\[4pt]
    \omega'(\lambda)
    &\;\xrightarrow{\;\epsilon\to 0\;}
    \sqrt{1+k'^{2}}
    = \omega_{0}\cosh\lambda + k_{0}\sinh\lambda,
    \label{eq:boost_limit_omega}
\end{align}
recovering the standard Lorentz boost.

The finite boost acting on the physical phonon field $\psi$ is obtained by composing the auxiliary-field boost~\eqref{eq:finiteboost_k}-\eqref{eq:finiteboost_omega} with the action of the unitary map $U$ in Fourier space. Since $\psi = U\phi$ and $U$ acts in Fourier space as multiplication by $e^{-\frac{i}{2}\epsilon T k^{2}}$, a mode of $\phi$ oscillating at auxiliary frequency $\omega_{\phi}(k)$ becomes a mode of $\psi$ oscillating at the shifted physical frequency
\begin{equation}
    \omega_{\psi}(k)
    = \omega_{\phi}(k) + \frac{\epsilon}{2}\,k^{2}
    = \sqrt{1 + k^{2} + \frac{\epsilon^{2}}{4}k^{4}}
      + \frac{\epsilon}{2}\,k^{2} \, .
\label{eq:freq_shift}
\end{equation}
The boosted wavenumber $k'_{a}(\lambda)$ is thus given by exactly the same formula~\eqref{eq:finiteboost_k} as for the auxiliary field, while the boosted physical frequency is obtained by applying the frequency shift~\eqref{eq:freq_shift} to the boosted auxiliary frequency evaluated at the boosted wavenumber $k'_{a}(\lambda)$:
\begin{align}
    k'_{a}(\lambda)
    &= \sqrt{\alpha_{-}}\;
       \mathrm{sc}(u_{0} + \mu\lambda,\;\kappa),
    \label{eq:finiteboost_psi_k} \\[4pt]
    \omega'_{\psi}(\lambda)
    &= \sqrt{1 + k_{a}^{\prime\,2}
             + \frac{\epsilon^{2}}{4}k_{a}^{\prime\,4}}
      + \frac{\epsilon}{2}\,k_{a}^{\prime\,2} \, .
    \label{eq:finiteboost_psi_omega}
\end{align}

It is also useful to characterize the finite boost directly in real space. In one spatial dimension, the deformed boost generator~\eqref{eq9} can be decomposed as $D=K+G$, where $K=T\partial_x+x\partial_T$ is the ordinary Lorentz boost generator and $G=-(\epsilon^2/2)T\partial_x^3$ is the higher-derivative deformation. The flow generated by $K$ acts geometrically on the spacetime coordinates according to
\begin{equation}
T'=T\cosh\lambda+x\sinh\lambda\,,
\qquad
x'=x\cosh\lambda+T\sinh\lambda\,.
\label{eq:realspace_geom}
\end{equation}
Thus, the geometric part of the symmetry remains the standard Lorentz transformation, while the deformation enters through the action of $G$ on the field configuration.

The exact finite transformation associated with the one-parameter flow generated by~\eqref{eq9} is obtained by exponentiating the complete generator,
\begin{equation}
\phi_\lambda=e^{\lambda D}\phi\,,
\qquad
\phi_{\lambda=0}=\phi\,.
\label{eq:realspace_field}
\end{equation}
Equivalently, $\phi_\lambda$ satisfies the boost-flow equation $d\phi_\lambda/d\lambda=D\phi_\lambda$. Since the geometric and higher-derivative contributions do not commute, $[K,G]\neq0$, the exact finite transformation cannot, in general, be factorized into an ordinary Lorentz transformation followed by an independent exponential of $G$. 

A convenient exact representation is obtained by passing to the interaction picture with respect to $K$. Writing $e^{\lambda D}=e^{\lambda K}\mathcal{U}_I(\lambda)$ and defining $G_I(\lambda)=e^{-\lambda K}Ge^{\lambda K}$, one obtains
\begin{equation}
e^{\lambda D}
=
e^{\lambda K}
\mathcal{P}
\exp\!\left[
\int_0^\lambda ds\,
e^{-sK}Ge^{sK}
\right]\,,
\label{eq:realspace_ordered}
\end{equation}
where $\mathcal{P}$ denotes ordering with respect to the boost parameter $s$. This ordering is required because the transported deformation operators $G_I(s)$ evaluated at different values of $s$ do not, in general, commute.

The conjugated operator appearing in~\eqref{eq:realspace_ordered} can be evaluated explicitly. The Lorentz flow gives $e^{-sK}Te^{sK}=T\cosh s-x\sinh s$ and $e^{-sK}xe^{sK}=x\cosh s-T\sinh s$, while the derivatives transform as $e^{-sK}\partial_xe^{sK}=\partial_x\cosh s+\partial_T\sinh s$ and $e^{-sK}\partial_Te^{sK}=\partial_T\cosh s+\partial_x\sinh s$. It follows that
\begin{equation}
G_I(s)
=
-\frac{\epsilon^2}{2}
\left(T\cosh s-x\sinh s\right)
\left(\partial_x\cosh s+\partial_T\sinh s\right)^3\,.
\label{eq:GI_explicit}
\end{equation}
The expression~\eqref{eq:GI_explicit} makes explicit why the finite deformation cannot be represented solely by an operator proportional to $T(s)\partial_x^3$. Under the Lorentz flow, the spatial derivative itself transforms and therefore generates mixed temporal and spatial derivative operators.

Expanding the cubic differential operator in~\eqref{eq:GI_explicit} gives
\begin{align}
G_I(s)
=
-\frac{\epsilon^2}{2}
\left(T\cosh s-x\sinh s\right)
\Big[
&\cosh^3 s\,\partial_x^3
+3\cosh^2 s\,\sinh s\,\partial_x^2\partial_T
\nonumber\\
&+3\cosh s\,\sinh^2 s\,\partial_x\partial_T^2
+\sinh^3 s\,\partial_T^3
\Big]\,.
\label{eq:GI_expanded}
\end{align}
Thus, although the infinitesimal generator~\eqref{eq9} remains a local, finite-order differential operator, its exact finite action is naturally represented by an ordered operator exponential. This is a consequence of exponentiating noncommuting differential operators and does not imply any loss of locality at the level of the infinitesimal symmetry.

Since $G=O(\epsilon^2)$, the exact representation~\eqref{eq:realspace_ordered} also provides a systematic perturbative expansion. To leading nontrivial order,
\begin{equation}
e^{\lambda D}
=
e^{\lambda K}
\left[
1+\int_0^\lambda ds\,G_I(s)
\right]
+O(\epsilon^4)\,.
\label{eq:realspace_leading}
\end{equation}
Accordingly, the action on the field becomes
\begin{equation}
\phi_\lambda
=
e^{\lambda K}
\left[
1
-\frac{\epsilon^2}{2}
\int_0^\lambda ds\,
\left(T\cosh s-x\sinh s\right)
\left(\partial_x\cosh s+\partial_T\sinh s\right)^3
\right]\phi
+O(\epsilon^4)\,.
\label{eq:realspace_leading_field}
\end{equation}
The expression~\eqref{eq:realspace_leading_field} therefore provides a controlled real-space representation of the finite transformation through $O(\epsilon^2)$ without factorizing the finite transformation into separate Lorentz and deformation operators.

The same reasoning extends to higher spatial dimensions. For a boost along the $a$-th spatial direction, one has $D_a=K_a+G_a$, where $K_a=T\partial_{x_a}+x_a\partial_T$ and $G_a=-(\epsilon^2/2)T\partial_{x_a}\nabla^2$. The exact finite transformation is represented by
\begin{equation}
e^{\lambda D_a}
=
e^{\lambda K_a}
\mathcal{P}
\exp\!\left[
\int_0^\lambda ds\,
e^{-sK_a}G_a e^{sK_a}
\right]\,.
\label{eq:realspace_higherd_ordered}
\end{equation}
In dimensions greater than one, the transported deformation acquires additional tensorial structure because both $\partial_{x_a}$ and $\nabla^2$ transform under the boost. Nevertheless, the underlying structure remains unchanged: the ordinary Lorentz generator determines the geometric coordinate transformation, while the higher-derivative contribution is incorporated through the ordered exponential of the Lorentz-transported deformation.

\section{Kinematical and spectral gamma factors}

We now turn to the observational consequences of our results, which may in principle be testable in experimental realizations of the viscoelastic fluid \eqref{eq2}, or in some superfluid systems \cite{silke-visser,ferrari22,eg2,marino2024}, given the correspondence between the phonon field $\psi$ and the Bogoliubov field $\phi$. A particularly relevant effect is the recently discussed separation of kinematical and spectral gamma factors in DSR models \cite{AmelinoCamelia2025TimeDilation}. These theories predict a breakdown of the usual identification between kinematical and dynamical quantities, leading to inequivalent gamma factors associated with proper time, energy, and wave propagation. These observations allow one, in principle, to distinguish between Lorentz invariance violation (LIV) scenarios, which assume undeformed transformation laws and therefore predict standard time dilation, and DSR scenarios, where deviations from the conventional relation $E / m$ are expected.

As clarified in \cite{AmelinoCamelia2025TimeDilation} a modified time-dilation factor indeed requires not only a modified dispersion relation, but also a consistent DSR-like spacetime sector including finite deformed boosts, an invariant notion of interval, and a consistent definition of proper time compatible with the deformed symmetry. Our hydrodynamic analogy provides precisely these ingredients within a concrete physical system, allowing us to implement a symmetry-based definition of time dilation and to compute the corresponding gamma factor explicitly.

Following the symmetry-based approach of Ref.~\cite{AmelinoCamelia2025TimeDilation}, we define time dilation from the transformation of the rest-frame proper time of a clock, and only subsequently express the result in terms of the spectral variables of the boosted mode.

We first work in the auxiliary-field representation, where the finite boost is known exactly. The positive-frequency rest mode of Eq. \eqref{bogo} is the mode with $k_0=0$ and $\omega_{\phi,0}=1$.
A localized auxiliary clock can therefore be modeled by a narrow wave packet centered at this point of the on-shell manifold $\phi_{\rm rest}(\tau,\xi)=A(\xi)e^{-i\tau}$, where $(\tau,\xi)$ are the coordinates of the clock rest frame. Since the finite boost constructed in Sec.~III acts on spacetime through the ordinary Lorentz transformations \eqref{eq:realspace_geom}, the coordinates $(\tau,\xi)$ are given by the inverse map
\begin{equation}
 \tau=T\cosh\lambda-x\sinh\lambda,
 \qquad
 \xi=x\cosh\lambda-T\sinh\lambda.
 \label{eq:td_inverse_boost}
\end{equation}
The rest clock is located at fixed spatial position, so its worldline in the rest frame is $\xi=0$.
Using Eq.~\eqref{eq:td_inverse_boost}, the image of this worldline in the laboratory coordinates is
\begin{equation}
 x=V_{\lambda}T \, ,
 \label{eq:td_worldline_lab}
\end{equation}
where $V_{\lambda}=\tanh\lambda$. Along this trajectory one has $dx=V_{\lambda}dT$, and Eq.~\eqref{eq:td_inverse_boost} gives
\begin{align}
 d\tau
 &=\cosh\lambda\,dT-\sinh\lambda\,dx
 \nonumber\\
 &=\bigl(\cosh\lambda-V_{\lambda}\sinh\lambda\bigr)dT
 \nonumber\\
 &=\frac{dT}{\cosh\lambda}.
 \label{eq:td_interval_from_boost}
\end{align}
Equivalently, since the coordinate sector is Lorentzian, the same relation follows from the invariant interval,
\begin{equation}
 d\tau^2=dT^2-dx^2
 =\bigl(1-V_{\lambda}^2\bigr)dT^2,
 \label{eq:td_interval_metric}
\end{equation}
which again implies
\begin{equation}
 dT=\gamma(\lambda)\,d\tau,
 \qquad
 \gamma(\lambda)=\cosh\lambda=\frac{1}{\sqrt{1-V_{\lambda}^{2}}}.
 \label{eq:td_gamma_lambda}
\end{equation}
Equation~\eqref{eq:td_gamma_lambda} is the proper-time dilation factor obtained directly from the finite boost. At fixed rapidity parameter $\lambda$ it is exactly Lorentzian. Thus the proper time-dilation law itself is not modified. The deformation enters when $\lambda$ is expressed in terms of the spectral variables of the boosted mode, since the orbit generated by the deformed boost in momentum space is no longer hyperbolic.

For a rest clock, the rapidity associated with a boosted mode of wavenumber $k$ is given by Eq. \eqref{eq:integral}. Substituting this relation into Eq.~\eqref{eq:td_gamma_lambda}, we obtain the kinematical proper-time factor parametrized by the boosted wavenumber for the auxiliary field $\phi$,
\begin{equation}
 \gamma_{\phi}(k)
 =\cosh\!\left[
 \frac{1}{\mu}
 F\!\left(\arctan\frac{k}{\sqrt{\alpha_-}},\kappa\right)
 \right].
 \label{eq:td_gamma_phi_k}
\end{equation}
The corresponding observer velocity is
\begin{equation}
 V_{\phi}(k)
 =\tanh\!\left[
 \frac{1}{\mu}
 F\!\left(\arctan\frac{k}{\sqrt{\alpha_-}},\kappa\right)
 \right].
 \label{eq:td_velocity_phi_k}
\end{equation}
This velocity is fixed by the spacetime action of the exact boost and should not be confused with the wavepacket group velocity,
\begin{equation}
 v_g(k)=\frac{d\omega_{\phi}}{dk}
 =\frac{k\left(1+\frac{\epsilon^2}{2}k^2\right)}{\sqrt{1+k^2+\frac{\epsilon^2}{4}k^4}},
 \label{eq:td_group_velocity}
\end{equation}
nor with the phase velocity $v_{\rm ph}(k)=\omega_{\phi}(k)/k$ for $k\neq0$. 

For $\epsilon=0$, when the dispersion relation reduces to the standard relativistic form, the boost parameter selects a velocity $V(k)=\tanh\lambda(k)$ that coincides with the group velocity, $v_g(k)=d\omega/dk$. This reflects the usual identification, in Lorentz-invariant theories, between the kinematical velocity defined by boosts and the propagation velocity of wave packets. In the further limit of a massless field, where the dispersion relation becomes linear, $\omega=|k|$, one also has $v_{\rm ph}=v_g$, so that all characteristic velocities coincide. 

For $\epsilon\neq0$, this correspondence breaks down: the boost-defined velocity $V_{\phi}(k) $ becomes a purely geometric quantity associated with the action of the deformed symmetry on spacetime, while $v_g$ and $v_{\rm ph}$ retain their dynamical meaning determined by the dispersion relation. In general, therefore, $V_{\phi}(k)\neq v_g(k)$ and $V_{\phi}(k)\neq v_{\rm ph}(k)$.
This mismatch highlights the separation between the kinematics encoded in the deformed symmetry and the dynamical propagation governed by the dispersion relation. 

At leading order in $\epsilon$ we find
\begin{equation}
 \gamma_{\phi}(k)
 =\sqrt{1+k^2}
 -\frac{\epsilon^2}{16\sqrt{1+k^2}}
 \left[
 k^4+3k^2-3k\sqrt{1+k^2}\,\operatorname{arsinh}k
 \right]
 +O(\epsilon^4) \, 
 \label{eq:td_gamma_phi_expansion}
\end{equation}
which can be written as
\begin{equation}
\gamma_{\phi}(k) \approx \gamma_0 \left(1 - \epsilon^2 \frac{g(k)}{16 \gamma_0^2}\right)
\end{equation}
where $\gamma_{0}=\sqrt{1+k^2}$ is the standard "gamma factor" for $\epsilon=0$ and $g(k)$ collects all terms appearing in the square parentheses on the rhs of Eq. \eqref{eq:td_gamma_phi_expansion}.

The unitary map $U$ relating the auxiliary Bogoliubov field to the physical phonon fluctuations does not act on the spacetime coordinates and therefore leaves the proper-time dilation factor in Eq.~\eqref{eq:td_gamma_lambda} unchanged. Hence, the boost-defined proper-time dilation factor obtained in the auxiliary representation is also the time-dilation factor associated with the physical phonon field $\psi$, which we denote by $\gamma_{\psi}$.

It is important, however, to distinguish this quantity from the frequency ratio associated with the physical phonon dispersion. In the present dimensionless variables,
\begin{equation}
 \widetilde\gamma_{\phi}(k)=\frac{\omega_{\phi}(k)}{\omega_{\phi}(0)}=\omega_{\phi}(k),
 \qquad
 \widetilde\gamma_{\psi}(k)=\frac{\omega_{\psi}(k)}{\omega_{\psi}(0)}=\omega_{\psi}(k),
 \label{eq:td_energy_gamma} \, .
\end{equation}
These quantities play the role of energy-over-rest-frequency factors. In the undeformed limit they coincide with the proper-time dilation factor, but for $\epsilon\neq0$ they are generically inequivalent:
\begin{equation}
 \gamma_{\phi}(k)\neq\widetilde\gamma_{\phi}(k),
 \qquad
 \gamma_{\psi}(k)\neq\widetilde\gamma_{\psi}(k).
 \label{eq:td_gamma_distinction}
\end{equation}
This inequivalence is the direct analogue, in the present hydrodynamic setting, of the separation between time-dilation and energy gamma factors emphasized in Ref.~\cite{AmelinoCamelia2025TimeDilation}. \\

\section{Discussion and conclusions}

In this work, we constructed an explicit realization of deformed Lorentz symmetry in a concrete hydrodynamic analogue system. By introducing a time-dependent transformation that diagonalizes the dynamics, we mapped the phonon field onto an auxiliary field obeying a modified Klein-Gordon equation with a purely spatial quartic
correction, for which the deformed symmetry admits a local, finite-order differential realization. This is not merely a technical result, but actually highlights an exceptional property of this modified Klein-Gordon field. In generic DSR-inspired field theories, deformations of the Lorentz algebra in momentum space typically lead, upon Fourier transformation, to symmetry generators that are nonlocal in position space. In this sense, quartic dispersion relations of the form \eqref{bogo} appear to lie at a special algebraic boundary between the local (undeformed) symmetry transformations of standard special relativity and the fully nonlocal transformations characteristic of DSR frameworks.

Remarkably, up to the rest-mass term, Eq.~\eqref{bogo} coincides with the Bogoliubov dispersion relation of linear excitations in Bose gases \cite{bogoliubov,reviewBEC}. Moreover, for homogeneous background flows, the Bogoliubov–de Gennes equations governing these fluctuations can be recast in the form of a Klein–Gordon equation with a quartic spatial derivative, as in Eq.~\eqref{eq7} (see, e.g., Eq.~(43) in \cite{prain} or Eq.~(17) in \cite{marino2019} for $\gamma=0$). Our results therefore establish a concrete and previously unrecognized connection between deformed relativistic kinematics and Bogoliubov-type excitations in condensed-matter systems.

Mapping back from the auxiliary Bogoliubov field $\phi$ to the physical phonon field $\psi$ via the inverse transformation $U^{-1}$ yields the corresponding boost generator $D_a^\psi = U D_a U^{-1}$, which, in spite of the nonlocal nature of the transformation $U$ remains a finite-order local differential operator.

We have then analyzed some observational consequences of a deformed symmetry in our system, focusing in particular, on the recently discussed emergence of inequivalent kinematical and spectral ”gamma factors” in DSR theories \cite{AmelinoCamelia2025TimeDilation}.
These observations are of particular relevance, as they could provide the signature to distinguish between Lorentz invariance violation (LIV) scenarios, which assume undeformed transformation laws and therefore predict standard time dilation, and DSR scenarios, where deviations from the conventional relation $E / m$ are expected.

As clarified in \cite{AmelinoCamelia2025TimeDilation} a modified time-dilation factor indeed requires not only a modified dispersion relation, but also a full DSR-like spacetime sector including finite deformed boosts, an invariant notion of interval, and a consistent definition of proper time compatible with the deformed symmetry. Our hydrodynamic system naturally provide these ingredients. As a result, we obtain an explicit realization of the separation between time-dilation and energy gamma factors emphasized in Ref.~\cite{AmelinoCamelia2025TimeDilation}. This inequivalence signals a breakdown of the standard identification between kinematical and dynamical observables, rendering the notion of velocity representation-dependent. In contrast with purely kinematical constructions, this separation arises here dynamically from the nontrivial relation between rapidity and spectral variables induced by the deformed boost. This provides a concrete example in which DSR-like phenomenology emerges without modifying the underlying spacetime metric.

Our hydrodynamic analogy, however, does not extend straightforwardly to multiparticle states, and this limitation highlights a deep structural feature of DSR theories. In the present hydrodynamic model, phonon momenta compose linearly: the total momentum of a two-phonon state is the ordinary sum of the individual momenta. A deformed boost, however, generically implies a nonlinear composition law: the transformed total momentum of a multiparticle state is not the sum of the individually transformed momenta unless the boost acts through a nontrivial coproduct on the tensor product of single-particle Hilbert spaces \cite{kowalskiPLB2002}. The present model provides no such coproduct: the deformed boost acts correctly on single-particle states but does not extend to a consistent action on multiparticle states that would simultaneously preserve both the deformed kinematics and the linear composition of momenta. 

This is not a peculiar feature of our analogue model. In DSR theories, the specification of a coproduct structure is an essential part of the physical content of the theory, and different choices lead to different predictions for multiparticle observables. By contrast, in condensed-matter systems the composition of momenta is fixed by translational invariance and is typically linear (additive), reflecting standard momentum conservation. The hydrodynamic model thus applies to the single-particle sector of the deformed kinematics, the sector that is best understood and most studied in existing phenomenological analyses, and realizes it exactly, while leaving the multiparticle sector in its standard, undeformed form. 

Whether a theory that realizes deformed Lorentz symmetry at the single-particle level should be classified as a genuine DSR model or as a Lorentz-violating theory is a question that cannot be settled by the dispersion relation alone. At a minimum, the former requires a consistent specification of observer transformations, which is precisely what the deformed boosts of Sec. III provide. These issues touch on general conceptual aspects of the relationship between DSR and LIV scenarios, which go beyond the scope of the present paper and will be the subject of a forthcoming work \cite{wip}.


\begin{thebibliography}{99}

\bibitem{amelino1} G. Amelino-Camelia, Testable scenario for Relativity with minimum-length. Phys. Lett. B, {\bf 510}, 255 (2001).
\bibitem{amelino2} G. Amelino-Camelia, Relativity in space-times with short-distance structure governed by an observer-independent (Planckian) length scale. Int. J. Mod. Phys. D, {\bf 11}, 35-59 (2002).
\bibitem{amelino3} G. Amelino-Camelia, Doubly-special relativity: first results and key open problems. J. Mod. Phys. D, {\bf 11}, 1643-1669 (2002).
\bibitem{kowalskiPLB2002} J.~Kowalski-Glikman and S.~Nowak,
"Doubly special relativity theories as different bases of kappa Poincare algebra", Phys. Lett. B \textbf{539} (2002), 126-132.
\bibitem{ms} J. Magueijo, and L. Smolin, Lorentz invariance with an invariant energy scale. Phys. Rev. Lett. {\bf 88}, 190403 (2002).
\bibitem{mattingly2005} D. Mattingly, Living Rev. Relativity {\bf 8}, 5 (2005).
\bibitem{liberati2013} S. Liberati, Classical Quantum Gravity {\bf 30}, 133001 (2013).
\bibitem{jafari} N. Jafari and B. Shukirgaliyev Nonrelativistic limits of the Klein-Gordon and Dirac equations in the Amelino-Camelia DSR. Phys. Lett. B {\bf 853}, 138693 (2024).
\bibitem{GACandMajid}
G.~Amelino-Camelia and S.~Majid,
``Waves on noncommutative space-time and gamma-ray bursts,''
Int. J. Mod. Phys. A \textbf{15} (2000), 4301-4324
[arXiv:hep-th/9907110 [hep-th]].

\bibitem{DimitrijevicJHEP2011}
M.~Dimitrijevic and L.~Jonke,
``A Twisted look on kappa-Minkowski: U(1) gauge theory,''
JHEP \textbf{12} (2011), 080
[arXiv:1107.3475 [hep-th]].

\bibitem{urrutia}
J.~Alfaro, H.~A.~Morales-Tecotl and L.~F.~Urrutia,
``Quantum gravity corrections to neutrino propagation,''
Phys. Rev. Lett. \textbf{84} (2000), 2318-2321
[arXiv:gr-qc/9909079 [gr-qc]].

\bibitem{hanno}
H.~Sahlmann and T.~Thiemann,
``Towards the QFT on curved space-time limit of QGR. 1. A General scheme,''
Class. Quant. Grav. \textbf{23} (2006), 867
[arXiv:gr-qc/0207030 [gr-qc]].

\bibitem{agrev} C. Barcel\'o, S. Liberati, and M. Visser, Analogue Gravity. Living Rev. Rel. {\bf 14}, 3 (2011).
\bibitem{natrev} S. L. Braunstein, M. Faizal, L. M. Krauss, F. Marino and N. A. Shah. Analogue simulations of quantum gravity with fluids. Nat. Rev. Phys. {\bf 5}, 612-622 (2023).
\bibitem{rousseaux} G. Rousseaux, The basics of water waves theory for analogue gravity, in Lecture Notes in Physics (Springer International Publishing, 2013) pp. 81-107.
\bibitem{bogoliubov} N. N. Bogoliubov, On the theory of superfluidity, J. Phys. (USSR) {\bf 11}, 23 (1947).
\bibitem{reviewBEC} F. Dalfovo, S. Giorgini, L. P. Pitaevskii, and S. Stringari, Theory of Bose-Einstein condensation in trapped gases, Rev. Mod. Phys. {\bf 71}, 463 (1999).
\bibitem{vocke15} D. Vocke, T. Roger, F. Marino, E. M. Wright, I. Carusotto, M. Clerici, and D. Faccio, "Experimental characterization of nonlocal photon fluids," Optica {\bf 2}, 484-490 (2015). 
\bibitem{fontaine18} Q. Fontaine, T. Bienaim\'e, S. Pigeon, E. Giacobino, A. Bramati, and Q. Glorieux, Observation of the Bogoliubov dispersion in a fluid of light. Phys. Rev. Lett. {\bf 121}, 183604 (2018).
\bibitem{falque} K. Falque, A. Delhom, Q. Glorieaux, E. Giacobino, A. Bramati, and M. J. Jacquet, Polariton Fluids as Quantum Field Theory Simulators on Tailored Curved Spacetimes. arXiv:2311.01392 (2023).
\bibitem{marino25} F. Marino, Phonons mimicking doubly special relativity kinematics, Phys. Lett. B {\bf 866}, 139529 (2025).
\bibitem{AmelinoCamelia2025TimeDilation}
G.~Amelino-Camelia, G.~Gubitosi, P.~Pellecchia, M.~Refuto, and G.~Rosati,
``Doubly special relativistic spacetime picture and the phenomenology of Planck-scale-modified time dilation,''
Phys.\ Rev.\ D \textbf{112}, 106006 (2025).
\bibitem{landau} L. D. Landau, E. M. Lifshitz, Fluid Mechanics (Vol. 6 of Course of Theoretical Physics) Pergamon Press (1987
\bibitem{chhabra} R. P. Chhabra, Non-newtonian fluids: an introduction. In Rheology of Complex Fluids (eds. Krishnan, J. M. Deshpande, A. P. Kumar, P. B. S.) (Springer, 2010).
\bibitem{irgens} F. Irgens, Rheology and Non-Newtonian Fluids, Springer, New York, 2014.
\bibitem{ferry} D. Ferry, Viscoelastic properties of polymers, 3rd ed. (John Wiley \& Sons, New York, NY, 1980).
\bibitem{winter} H. H. Winter and F. Chambon, "Analysis of linear viscoelasticity of a crosslinking polymer at the gel point," J. Rheol. {\bf 30}, 367-382 (1986).
\bibitem{marino2019} F. Marino, Massive phonons and gravitational dynamics in photon-fluids. Phys. Rev. A {\bf 100}, 063825 (2019).
\bibitem{ciszak2021} M. Ciszak and F. Marino, Acoustic black-hole bombs and scalar clouds in a photon-fluid model. Phys. Rev. D {\bf 103}, 045004 (2021).
\bibitem{marino2024} M. Ciszak and F. Marino, Bound Bogoliubov quasiparticles in photon superfluids. Phys. Rev. Research {\bf 6}, L032006 (2024).
\bibitem{silke-visser} M. Visser and S. Weinfurtner, Massive Klein-Gordon equation from a Bose-Einstein-condensation-based analogue spacetime. Phys. Rev. D {\bf 72}, 044020 (2005).
\bibitem{ferrari22} R. Cominotti, A. Berti, A. Farolfi, A. Zenesini, G. Lamporesi, I. Carusotto, A. Recati, and G. Ferrari. Observation of Massless and Massive Collective Excitations with Faraday Patterns in a Two-Component Superfluid. Phys. Rev. Lett. {\bf 128}, 210401 (2022).
\bibitem{eg2} F. Girelli, S. Liberati and L. Sindoni, Gravitational dynamics in Bose-Einstein condensates. Phys. Rev. D {\bf 78}, 084013 (2008).
\bibitem{prain} A. Prain, C. Maitland, D. Faccio, F. Marino, Phys. Rev. D {\bf 100}, 024037 (2019).
\bibitem{luki} J. Lukierski, A. Nowicki, H. Ruegg and V.N. Tolstoy, Phys. Lett. B {\bf 264},
331 (1992).
\bibitem{luki2} J. Lukierski and A. Nowicki, Int. J. Mod. Phys. A {\bf 18}, 7 (2003).
\bibitem{strogatz} S. H. Strogatz, Nonlinear Dynamics and Chaos (Westview Press, Boulder, USA, 2015).
\bibitem{wip} A. Shabir, F. Marino and M. Faizal, One-particle equivalence of Lorentz-violating and Lorentz-deformed theories, arXiv:2611.09986 (2026).

\end{thebibliography}
\end{document}